\documentclass[fleqn,10pt]{wlscirep}
\definecolor{color1}{rgb}{0,0,0}
\usepackage[utf8]{inputenc}
\usepackage[T1]{fontenc}
\usepackage{xspace}
\usepackage{longtable}
\usepackage{tikz}
\renewenvironment{abstract}{}{}

\usepackage[superscript,biblabel]{cite}
\newcommand{\aj}{The Astronomical Journal}
\newcommand{\apj}{The Astrophysical Journal}

\newcommand{\aap}{Astronomy \& Astrophysics}
\newcommand{\mnras}{Monthly Notices of the Royal Astronomical Society}
\newcommand{\apjs}{The Astrophysical Journal Supplement Series}

\newcommand{\icarus}{Icarus}
\newcommand{\planss}{Planetary and Space Science}

\newcommand{\tess}{TESS\,}

\newcommand{\mjup}{$M\rm_J$\,}
\newcommand{\rjup}{$R\rm_J$\,}
\newcommand{\mearth}{$M_\oplus$}

\newcommand{\planetb}{TOI-201\,b\,}
\newcommand{\planetc}{TOI-201\,c\,}
\newcommand{\planetd}{TOI-201\,d\,}
\newcommand{\figinteriormodels}{Extended Data Fig. 1}

\vspace{-2cm}
\title{A very eccentric brown dwarf coplanar to a warm Jupiter and a hot super Earth}
\vspace{-2cm}

\author[1,*]{Mat\'ias I. Jones}
\author[2]{Luca Naponiello}
\author[3,4]{Trifon Trifonov}
\author[5,6]{Rafael Brahm}
\author[7]{Gabriele Pichierri}
\author[8]{Lorena Acu\~{n}a-Aguirre}
\author[1]{Robert J. De Rosa}
\author[5,6]{Marcelo Tala Pinto}
\author[2]{Aldo S. Bonomo}
\author[9,2,8]{Luigi Mancini}
\author[2]{Alessandro Sozzetti}
\author[8]{Yared Reinarz}
\author[10]{Alessandro Morbidelli}
\author[11]{N\'estor Espinoza}
\author[7]{Giovanni Rosotti}
\author[12]{Eric L. Nielsen}
\author[4,13]{Stefan Y. Stefanov}
\author[8]{Thomas Henning}
\author[5,6]{Andr\'es Jord\'an}
\author[8]{Jan Eberhardt}
\author[14]{Artie Hatzes}
\author[15,16]{Leonardo Vanzi}
\author[17]{Jan Janik}
\author[18]{Petr Kabath}

\affil[1]{European Southern Observatory, Alonso de C\'ordova 3107, Vitacura, Casilla 19001, Santiago, Chile}
\affil[2]{INAF -- Osservatorio Astrofisico di Torino, Via Osservatorio 20, 10025 Pino Torinese, Italy}
\affil[3]{Landessternwarte, Zentrum f\"ur Astronomie der Universt\"at Heidelberg, K\"onigstuhl 12, 69117 Heidelberg, Germany}
\affil[4]{Department of Astronomy, Faculty of Physics, Sofia University ``St Kliment Ohridski'', 5 James Bourchier Blvd, 1164 Sofia, Bulgaria}
\affil[5]{Facultad de Ingenier\'ia y Ciencias, Universidad Adolfo Ib\'a\~nez, Av.\ Diagonal las Torres 2640, Pe\~nalol\'en, Santiago, Chile}
\affil[6]{Millennium Institute for Astrophysics, Camino El Observatorio 1515, Las Condes, Casilla, Santiago, Chile}
\affil[7]{Dipartimento di Fisica, Università degli Studi di Milano, Via Celoria 16, 20133 Milano, Italy}
\affil[8]{Max Planck Institute for Astronomy, K\"{o}nigstuhl 17, 69117 Heidelberg, Germany}
\affil[9]{Department of Physics, University of Rome ``Tor Vergata'', Via della Ricerca Scientifica 1, 00133 Rome, Italy}
\affil[10]{Collège de France, 11 Pl. Marcelin Berthelot, 75231 Paris, France}
\affil[11]{Space Telescope Science Institute, 3700 San Martin Drive, Baltimore, MD 21218, USA}
\affil[12]{Department of Astronomy, New Mexico State University, Las Cruces, NM 88003, USA}
\affil[13]{Institute of Astronomy and National Astronomical Observatory, Bulgarian Academy of Sciences, Tsarigradsko Shose 72, BG-1784 Sofia, Bulgaria}
\affil[14]{Th{\"u}ringer Landessternwarte, D-07778 Tautenburg, Germany}
\affil[15]{Center of Astro Engineering, Pontificia Universidad Cat\'olica de Chile, Av. Vicu\~na Mackenna 4860, 782-043 Santiago, Chile}
\affil[16]{Department of Electrical Engineering, Pontificia Universidad Católica de Chile, Av. Vicuña Mackenna 4860, 782-043 Santiago, Chile}
\affil[17]{Department of Theoretical Physics and Astrophysics, Faculty of Science, Masaryk University, Kotlarska 2, CZ-611 37, Brno, Czech Republic}
\affil[18]{Astronomical Institute, Czech Academy of Sciences, Fri\v{c}ova 298, 251 65 Ond\v{r}ejov, Czechia}
\affil[*]{Corresponding author: mjones@eso.org}

\date{}

\begin{document}
\maketitle
\noindent {\bf
In transiting planetary systems, where planetary sizes are accurately determined from transit observations,
the presence of transit timing variations (TTVs), especially when combined with radial velocity (RV) data,
provides powerful constraints on masses and orbital eccentricities. Together, these measurements offer crucial
insights into system architecture, formation mechanisms, and dynamical evolution. We present long-term RV
and transit/TTV monitoring of the active and young star (age $\sim$1 Gyr) TOI-201, revealing an exceptional
multi-planet system composed of a hot super-Earth (SE) transiting every 5.8 days, a warm Jupiter (WJ) on
a 53-day orbit, and an eccentric (e = 0.622) low-mass brown dwarf (BD) on an approximately 8-year orbit,
with an estimated mass of M$_{\rm BD}$ $\sim$ 16 Jupiter masses. The BD is the longest-period transiting object ever
characterized via RVs, and the only one known to be coplanar with inner planets. The architecture of this
system suggests that the SE was formed isolated and in the innermost region of the gaseous disc. On the other
hand, the orbital configuration of the outer companions suggests a nearly in-situ formation of both objects,
with the WJ forming in a dense inner disc. Alternatively, the BD might have formed farther out and migrated
inward, while inflating its eccentricity due to interactions with the disc.
}\newline \newline

\noindent TOI-201 is a young (age $\sim$ 1 Gyr) F-type main-sequence star with a V -band optical brightness of 9.1 magnitudes,
located 112 pc from the Sun (see Methods section ‘Stellar parameters’). This star is known to harbor a transiting
warm Jupiter (WJ; TOI-201 b ) with a 53-day orbital period, which was discovered in 2019 by the Transiting
Exoplanets Survey Satellite1 (TESS \cite{Ricker2015}). The planet was initially characterized using ground-based spectroscopic data
and was one of the first so-called WJs detected by \tess\cite{Hobson2021}. TOI-201 is also among the few systems that were observed in more than 30 TESS sectors, 17 of which were not yet available in the discovery paper (refer to the Methods
section ‘TESS photometric data’). Building on this expanded light curve dataset, we are now able not only to refine
the physical and orbital parameters of TOI-201 b , but also to validate a new 5.8-day periodic transit signal induced
by an inner super-Earth (SE), hereafter named TOI-201 c (refer to Methods section ‘super-Earth confirmation’).
Furthermore, a single long-duration transit event was detected in Sector 64 (April 2023). Ground-based spectroscopic
observations (see Methods section ‘High-resolution spectroscopic data’) in the context of the Warm gIaNts with
tEss (WINE) collaboration\cite{Brahm2019}, along with archival radial velocity (RV) data, reveal that this mono-transit event is due
to an eccentric long-period Brown Dwarf (BD), TOI-201 d . This is consistent both with SPHERE high-contrast
imaging observations, which fully rule out the presence of stellar binary companions within $\sim$ 5 arcsec (see details
in Methods section ‘VLT/SPHERE high-contrast imaging’), and with transit time variations (TTVs) observed for the
WJ. 

\begin{figure}[t]
\centering
\includegraphics[width=0.95\linewidth]{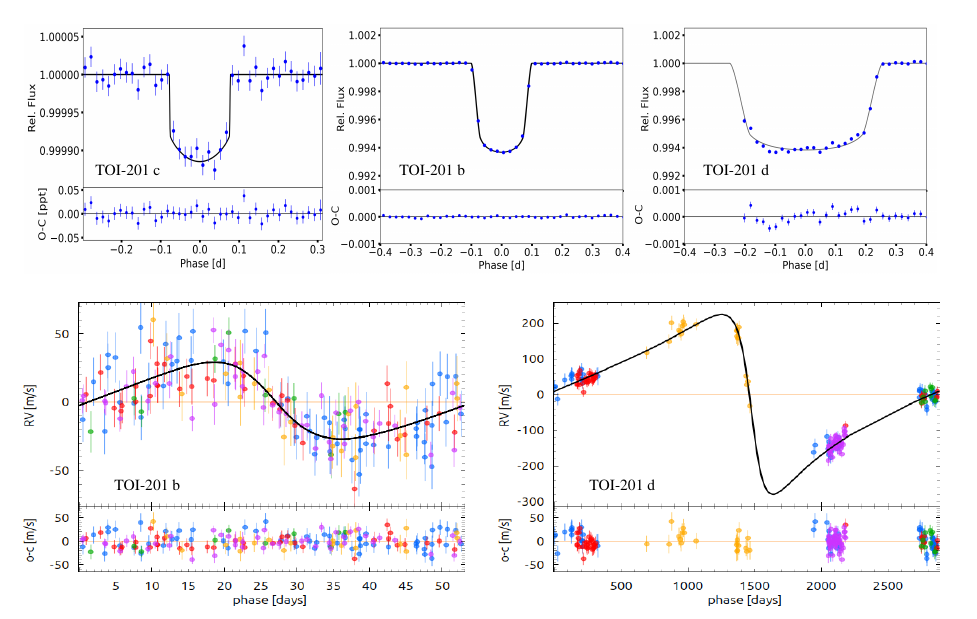}
\caption{Upper row: Phase folded transits of TOI-201 c, as obtained with 34 different sectors (left panel),
TOI-201 b , including 16 transits (middle panel), and the mono-transit event of TOI-201 d , in Sector 64 (right panel).
The blue dots correspond to 20-minute phase-bins, and the black solid line to the best transit model. Lower row:
Phase folded RV curve of TOI-201, as induced by planets b and d (left and right panels, respectively). The blue, red,
green, orange, and magenta dots correspond to FEROS, HARPS, CORALIE, CHIRON and PLATOSPEC data,
respectively. The solid line corresponds to the best-fitting model.\label{fig:trasit_RV_phase}}\end{figure}
\begin{table}\caption{Physical and orbital parameters of the inner planets and the BD.}  
\label{tab:planet_pars}   
\centering 
\resizebox{0.6\textwidth}{!}{
\begin{tabular}{l r r r}  
\hline\hline  
Parameter & TOI-201\,d & TOI-201\,b&  TOI-201\,c \\  
\hline\vspace{-0.4cm} \\ \vspace{0.1cm}
    Period (d)                  & 5.849240$^{+0.000005}_{-0.000006}$    & 52.97804$^{+0.00004}_{-0.00002}$ &2881.15$^{+0.02}_{-0.01}$\\ \vspace{0.1cm} 
    Semi-major axis (au)        & 0.0677$^{+0.0012}_{-0.0012}$          & 0.3001$^{+0.0031}_{-0.0031}$     & 4.33$^{+0.06}_{-0.06}$ \\ \vspace{0.1cm} 
    Radius                      & 1.436$^{+0.045}_{-0.046}$\,$R_\oplus$ & 1.01$^{+0.03}_{-0.03}$\,\rjup    & 0.99$^{+0.03}_{-0.03}$\,\rjup \\ \vspace{0.1cm} 
    Eccentricity                & 0.116$^{+0.154}_{-0.084}$             & 0.275$^{+0.002}_{-0.002}$        & 0.622$^{+0.001}_{-0.001}$\\ \vspace{0.1cm} 
    Mass                        & $<$ 33 \mearth                        &  0.59$^{+0.02}_{-0.02}$\,\mjup   & 16.46$^{+0.48}_{-0.48}$\,\mjup \\ \vspace{0.1cm} 
    Mean density (g\,cm$^{-3}$) &   $<$ 61                              &  0.71$\pm$0.07                   & 21.07$\pm$1.78        \\ \vspace{0.1cm}   
    Eq. temperature (K)         & 1337.6 $\pm$ 15.2                     & 635.7 $\pm$ 10.4                 & 167.8 $\pm$ 3.0 \\
    Orbital inclination (deg)   & 87.54$^{+0.68}_{-0.39}$ & 88.855$^{+0.001}_{-0.003}$ & 89.855$^{+0.001}_{-0.001}$  \vspace{-0.3cm} \\ \\ 
\hline     \hline  
\end{tabular}}
\end{table}
We performed a joint analysis of TESS photometry and high-precision RVs from five different facilities (see details in Methods section `High-resolution spectroscopy') using a self-consistent $N$-body model, which takes into account the mutual gravitational interaction between the WJ and the BD. This was necessary since the BD induces notable TTVs on the WJ (refer to the Methods section `Joint RV and transit $N$-body orbital fitting analysis'). 
The inner SE was modeled independently using a Keplerian orbital fit to the TESS light curves, as it is sufficiently distant from the massive companions to avoid detectable dynamical interactions on observable timescales. Moreover, its RV signal remains below the level of the observed stochastic RV noise (usually referred to as `stellar jitter'). As a result, we could determine its radius with high precision, but only place an upper limit on its mass of $\sim$ 33 Earth masses (\mearth) at the $1\sigma$ significance level (see Methods section `super-Earth confirmation'). 
Figure \ref{fig:trasit_RV_phase} shows all the phase-folded transits and RVs, along with their best-fitting models, while the main properties obtained for the planets and the BD are listed in Table \ref{tab:planet_pars}. The radius of \planetc is well below the radius valley that separates the population of SEs, small rocky planets, from sub-Neptunes, which are expected to retain a significant fraction of water or gas in their envelopes\cite{Rogers2015}. 
We compared the resulting planet radius with interior models with different masses and compositions (see Methods section `Interior models'), finding that \planetc is consistent with being rock-dominated, with no significant amount of water on its surface, and with a mass between $\sim$ 3\,\mearth, for a pure silicate rock composition, and $\sim 9$ \mearth\, for $100\%$ iron composition (\figinteriormodels).

\begin{figure}[!b]
\includegraphics[width=0.95\textwidth,angle=0]{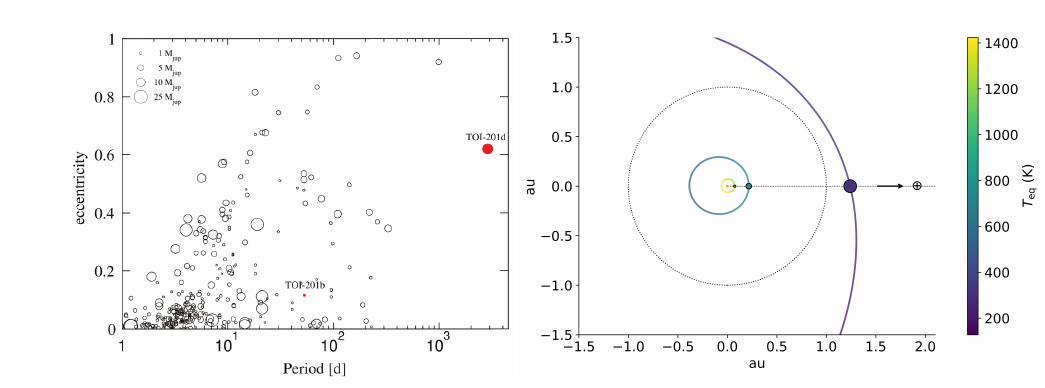}
\caption{{\it Left panel:} Eccentricity vs. orbital period for known transiting giants (0.3 $<$ $M_{\rm p}$/\mjup $<$ 30) with eccentricity and mass precision better than 20\%, including \planetb and \planetd (red circles).  For visual clarity, the symbol sizes scale with the square root of the planet mass. Data retrieved from the \href{https://exoplanetarchive.ipac.caltech.edu/}{NASA Exoplanet Archive} (as of May 2025), and complemented with the low-mass BD compilation in ref.\cite{Carmichael2023}
{\it Right panel}: Orbital configuration of the TOI-201 system. The solid lines represent the orbit of the three planets, with the color scale corresponding to different equilibrium temperature. For comparison a circular orbit at 1 AU (Earth) is also shown (dotted line).
\label{fig:per_ecc}}
\end{figure}

For \planetb, we confirmed that it is a prototypical WJ (see Fig.~\ref{fig:per_ecc} and Extended Data Fig. 2) in a mildly eccentric orbit ($e=0.280\pm0.002$), and we refined its orbital parameters. By comparing its measured radius with different MESA\cite{paxton2011} models (see Methods section `Interior models'), we found that \planetb is consistent with a rocky core with a mass between $\sim 14-28$\,\mearth, surrounded by a gaseous envelope with the same metal fraction as the parent star (Extended Data Fig. 3). 
This corresponds to a planet metallicity of $Z_{\rm p}=0.13^{+0.04}_{-0.03}$, thus to a planet heavy-element enrichment of $Z_{\rm p}/Z_\star$ = 5.0$^{+1.5}_{-1.2}$. These values are consistent with those obtained with GASTLI\cite{gastli} of $Z_{\rm p}=0.08^{+0.08}_{-0.05}$ and $Z_{\rm p}/Z_\star = 3.1^{+3.1}_{-1.9}$.
On the other hand, for \planetd we obtained $M_{\rm p} = 16.46^{+0.34}_{-0.34}$\,Jupiter masses (\mjup) and $R_{\rm p}=0.99^{+0.03}_{-0.03}$\,Jupiter radii (\rjup), placing this object within the BD mass regime, slightly above the deuterium-burning limit\cite{spiegel2011}, and consistent with models with low initial deuterium fraction (Extended Data Fig. 3). 
In particular, \planetd\, is the farthest transiting substellar companion ($P=2881.15^{+0.02}_{-0.01}$\,days and $e=0.622^{+0.001}_{-0.001}$) ever characterized with RV-follow up (see Figure \ref{fig:per_ecc}). TOI-201 is also the only known planetary system hosting a distant BD which is coplanar to inner planets, in this case a WJ and a SE.

\begin{figure}[ht!]
\centering
\includegraphics[width=0.75\linewidth]{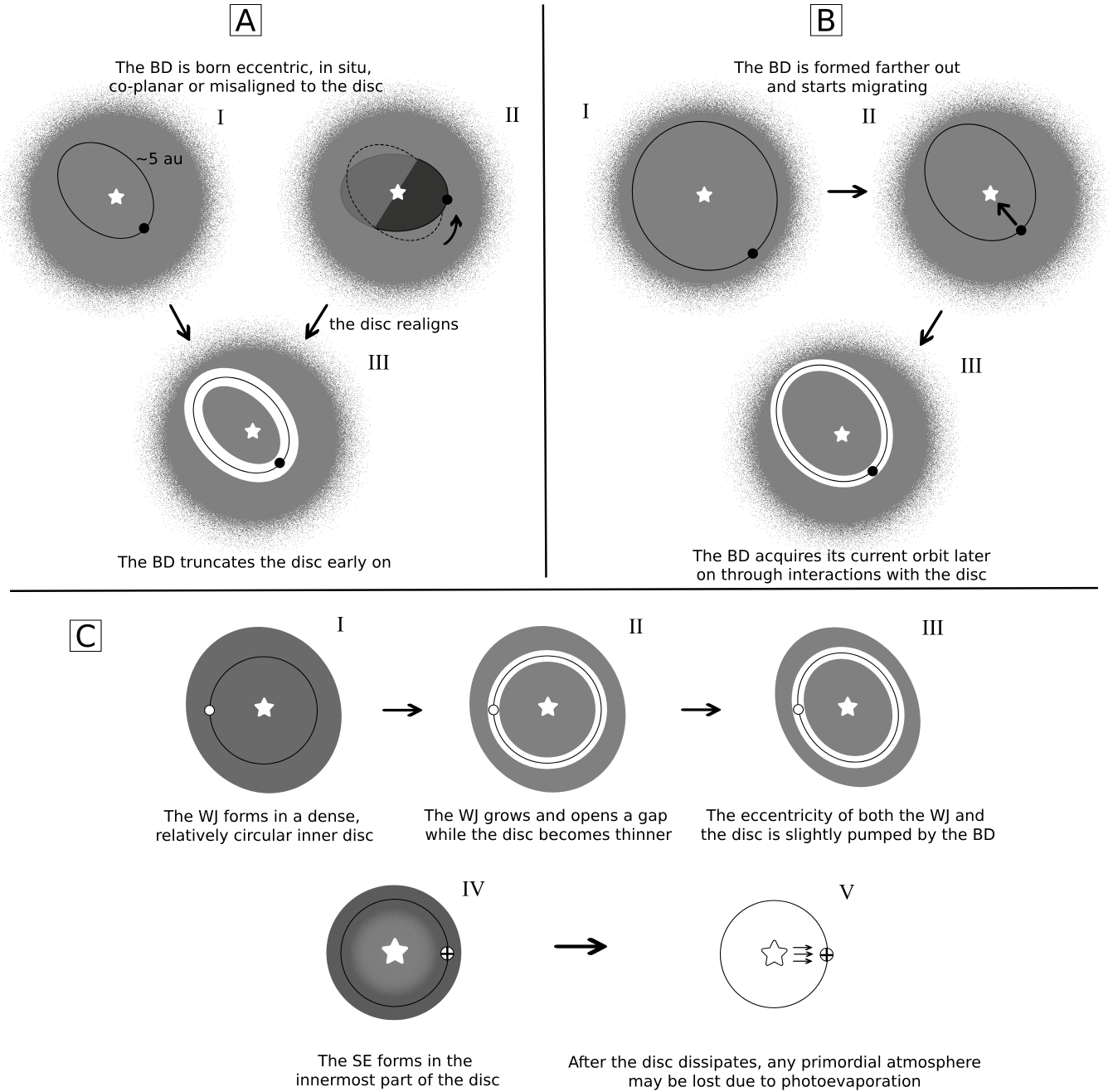}
\caption{Diagram of the possible formation scenarios discussed in the main text. Panel C focuses on the inner region where TOI-201\,b and TOI-201\,c form, with the enlarged star symbol denoting a smaller spatial scale.}\label{fig:formation_scenarios}
\end{figure}

The orbital configuration of the TOI-201 planetary system is shown in Figure \ref{fig:per_ecc}. Similarly, the current state of the WJ+BD system is depicted in Extended Data Figure 4 (a), where we note that the WJ has a rather large free eccentricity compared to the forced eccentricity, $e_\mathrm{WJ,free} > e_\mathrm{WJ,forced}$. Here, $e_\mathrm{WJ,forced}$ is the secular equilibrium eccentricity due to the BD, while $e_\mathrm{WJ,free}$ is the amplitude of the oscillating mode around the forced eccentricity\cite{Murray1999}. We note that instabilities and chaotic events such as planet-planet scattering\cite{Chatterjee2008} would most likely have generated significant mutual inclinations\cite{Juric2008}, which means that any instability is not expected to have taken place after the formation of the system, due to its co-planarity. This implies that the eccentricity of the WJ was acquired through less violent mechanisms. We can envision two possibilities for the assembly of this system (see Figure \ref{fig:formation_scenarios}). 
One is that the BD formed close to its current orbit early on, either as a fragment in a massive disc\cite{2007MNRAS.382L..30S} or alongside multiple fragments of which it was the only survivor, thus acquiring its high eccentricity \cite{2018MNRAS.474.5036F} (Panel A, I). Had it also acquired some mutual inclination with the remaining disc inside its orbit, the latter must have realigned with the BD\cite{Papaloizou1995, Larwood1996 ,Fragner2010} to ensure co-planarity between the BD, the WJ and the inner SE (Panel A, II). 
Such eccentric massive companion would truncate \cite{Artymowicz1994,Pichardo2005} and excite an eccentricity in the disc\cite{Kley2008} (Panel A, III), hindering planet formation since the material budget would be limited and the growth in an eccentric disc 
can be challenging\cite{2007ApJ...654..641J,2006Icar..183..193T,2008MNRAS.388.1528T,Muller2012,2013ApJ...765L...8R}. Both these issues point to an initially dense inner disc with a large enough solid budget and whose self-gravity drives a fast pericenter precession to decouple the disc-BD secular interactions, allowing it to remain (initially) relatively circular \cite{Marzari2009}.
The other possibility is that the BD formed farther out and even less eccentric (Panel B, I) , and that it migrated inward while getting its eccentricity pumped (Panel B, II), due to interactions with the gas\cite{2019MNRAS.483.2347D} (an eccentricity pumping of the BD via Kozai-Lidov cycles\cite{Kozai1962,Lidov1962} induced by a close companion is excluded by the SPHERE observations; see Extended Data Fig. 4), and reaching its current state (Panel B, III). 
This scenario would alleviate the issues posed by a severely truncated and eccentric disc, thus leaving more flexibility for the growth of the planets. However, we note that the WJ's orbit is rather close to the instability limit (see Extended Data Fig. 4), meaning that, for this scenario to work, the BD would have to arbitrarily stop its inward migration, so as not to disrupt the WJ. Thus, although possible, this explanation may be more fortuitous. A more distant BD, if it formed inclined with respect to the disc, would also require more time to allow for disc-BD realignment. \newline \indent
In any case, the WJ most likely formed in an initially dense, relatively circular disc (Panel C, I), with initially e$_{\rm WJ}$cos($\Delta\omega$)~$\sim$~e$_{\rm WJ}$ sin($\Delta\omega$)~ $\sim$~0, where $\Delta\omega$ = $\omega_{\rm BD}$ - $\omega_{\rm WJ}$ is the relevant secular angle (see Methods section ‘Orbital state and stability analysis’). 
Looking at Extended Data Fig. 4, this configuration means that e$_{\rm WJ,free}$ $\sim$ e$_{\rm WJ,forced}$, but the WJ is initially decoupled from the BD. 
Then, as the WJ became massive enough to open a gap\cite{1986ApJ...309..846L} (Fig. 3, Panel C, II), or as the disc became less dense and thus eccentric (or both), the eccentricity of the WJ could have been pumped\cite{Papaloizou2001,Kley2006,Bitsch2013,Ragusa2018}. The WJ started secularly evolving with the BD around e$_{\rm WJ,forced}$, and reached its currently observed state (Panel C, III) with  e$_{\rm WJ,free}$ $>$ e$_{\rm WJ,forced}$ due to e-pumping.

The inner SE also most likely formed in the disc (Panel C, IV), as gas-free formation\cite{2018MNRAS.479.5303L} is unlikely because of the low isolation mass at these locations. Moreover, the forced eccentricity near the current location of the SE is $\sim 0.1$; left-over planetary cores and planetesimals would experience inefficient growth due to high-velocity dispersion, with limited gravitational focusing and collisions leading to fragmentation/erosion rather than growth.

For the bulk of the SE population, the dichotomy\cite{2011ApJS..197....8L} between single- and multi-transiting systems can be explained by the assembly of multiple SEs within the same system and invoking post-disc instabilities \cite{2017MNRAS.470.1750I,2021A&A...650A.152I,2022ApJ...939L..19I}, even when outer giants are present \cite{2023A&A...674A.178B}. \newline 
In TOI-201, however, this scenario is disfavoured because the SE would have likely acquired some inclination with respect to the outer bodies, as well as a larger eccentricity after the instability compared to its current eccentricity, which is instead close to the forced eccentricity driven by the outer bodies. Unless additional yet-undetected planets
are orbiting the star on inclined orbits, this suggests that the SE may have formed alone in the innermost region of the disc. 
There, it could have acquired a primordial atmosphere which, given its current radius ($\sim 1.4 \, R_\oplus$), may have been removed (Panel C, V). Given the relatively young age of the system ($\sim$ 1 Gyr), photoevaporation\cite{Owen2013} appears to be a more viable atmospheric escape mechanism than core-powered mass loss\cite{Ginzburg2018}. 

The TOI-201 system represents an exceptional and unique multi-planet system, offering a rare window into the processes of planet formation and dynamical evolution. Its rare configuration provides invaluable insights into the formation and evolution pathways, suggesting a dynamically mild, coplanar and almost in-situ formation of a WJ, a highly eccentric BD, and a lone inner SE, within a common protoplanetary disk. Future studies, such as
high-cadence spectroscopic follow-up to measure the mass of the SE and the spin-orbit alignment of the WJ, will provide further information about the current architecture and the past evolution of this system.


\begin{thebibliography}{10}
\expandafter\ifx\csname url\endcsname\relax
  \def\url#1{\texttt{#1}}\fi
\expandafter\ifx\csname urlprefix\endcsname\relax\def\urlprefix{URL }\fi
\expandafter\ifx\csname doiprefix\endcsname\relax\def\doiprefix{DOI }\fi
\providecommand{\bibinfo}[2]{#2}
\providecommand{\eprint}[2][]{\url{#2}}

\bibitem{Ricker2015}
\bibinfo{author}{{Ricker}, G.~R.} \emph{et~al.}
\newblock \bibinfo{journal}{\bibinfo{title}{{Transiting Exoplanet Survey Satellite (TESS)}}}.
\newblock {\emph{\JournalTitle{Journal of Astronomical Telescopes, Instruments, and Systems}}} \textbf{\bibinfo{volume}{1}}, \bibinfo{pages}{014003} (\bibinfo{year}{2015}).

\bibitem{Hobson2021}
\bibinfo{author}{{Hobson}, M.~J.} \emph{et~al.}
\newblock \bibinfo{journal}{\bibinfo{title}{{A Transiting Warm Giant Planet around the Young Active Star TOI-201}}}.
\newblock {\emph{\JournalTitle{The Astronomical Journal}}} \textbf{\bibinfo{volume}{161}}, \bibinfo{pages}{235} (\bibinfo{year}{2021}).

\bibitem{Brahm2019}
\bibinfo{author}{{Brahm}, R.} \emph{et~al.}
\newblock \bibinfo{journal}{\bibinfo{title}{{HD 1397b: A Transiting Warm Giant Planet Orbiting A V = 7.8 mag Subgiant Star Discovered by TESS}}}.
\newblock {\emph{\JournalTitle{\aj}}} \textbf{\bibinfo{volume}{158}}, \bibinfo{pages}{45} (\bibinfo{year}{2019}).

\bibitem{Rogers2015}
\bibinfo{author}{{Rogers}, L.~A.}
\newblock \bibinfo{journal}{\bibinfo{title}{{Most 1.6 Earth-radius Planets are Not Rocky}}}.
\newblock {\emph{\JournalTitle{\apj}}} \textbf{\bibinfo{volume}{801}}, \bibinfo{pages}{41} (\bibinfo{year}{2015}).

\bibitem{paxton2011}
\bibinfo{author}{{Paxton}, B.} \emph{et~al.}
\newblock \bibinfo{journal}{\bibinfo{title}{{Modules for Experiments in Stellar Astrophysics (MESA)}}}.
\newblock {\emph{\JournalTitle{\apjs}}} \textbf{\bibinfo{volume}{192}}, \bibinfo{pages}{3} (\bibinfo{year}{2011}).

\bibitem{gastli}
\bibinfo{author}{{Acu{\~n}a}, L.}, \bibinfo{author}{{Kreidberg}, L.}, \bibinfo{author}{{Zhai}, M.} \& \bibinfo{author}{{Molli{\`e}re}, P.}
\newblock \bibinfo{journal}{\bibinfo{title}{{GASTLI. An open-source coupled interior-atmosphere model to unveil gas-giant composition}}}.
\newblock {\emph{\JournalTitle{\aap}}} \textbf{\bibinfo{volume}{688}}, \bibinfo{pages}{A60} (\bibinfo{year}{2024}).

\bibitem{spiegel2011}
\bibinfo{author}{{Spiegel}, D.~S.}, \bibinfo{author}{{Burrows}, A.} \& \bibinfo{author}{{Milsom}, J.~A.}
\newblock \bibinfo{journal}{\bibinfo{title}{{The Deuterium-burning Mass Limit for Brown Dwarfs and Giant Planets}}}.
\newblock {\emph{\JournalTitle{\apj}}} \textbf{\bibinfo{volume}{727}}, \bibinfo{pages}{57} (\bibinfo{year}{2011}).

\bibitem{Carmichael2023}
\bibinfo{author}{{Carmichael}, T.~W.}
\newblock \bibinfo{journal}{\bibinfo{title}{{Improved radius determinations for the transiting brown dwarf population in the era of Gaia and TESS}}}.
\newblock {\emph{\JournalTitle{\mnras}}} \textbf{\bibinfo{volume}{519}}, \bibinfo{pages}{5177--5190} (\bibinfo{year}{2023}).

\bibitem{Murray1999}
\bibinfo{author}{{Murray}, C.~D.} \& \bibinfo{author}{{Dermott}, S.~F.}
\newblock \emph{\bibinfo{title}{{Solar System Dynamics}}} (\bibinfo{year}{1999}).

\bibitem{Chatterjee2008}
\bibinfo{author}{{Chatterjee}, S.}, \bibinfo{author}{{Ford}, E.~B.}, \bibinfo{author}{{Matsumura}, S.} \& \bibinfo{author}{{Rasio}, F.~A.}
\newblock \bibinfo{journal}{\bibinfo{title}{{Dynamical Outcomes of Planet-Planet Scattering}}}.
\newblock {\emph{\JournalTitle{\apj}}} \textbf{\bibinfo{volume}{686}}, \bibinfo{pages}{580--602} (\bibinfo{year}{2008}).

\bibitem{Juric2008}
\bibinfo{author}{{Juri{\'c}}, M.} \& \bibinfo{author}{{Tremaine}, S.}
\newblock \bibinfo{journal}{\bibinfo{title}{{Dynamical Origin of Extrasolar Planet Eccentricity Distribution}}}.
\newblock {\emph{\JournalTitle{\apj}}} \textbf{\bibinfo{volume}{686}}, \bibinfo{pages}{603--620} (\bibinfo{year}{2008}).

\bibitem{2007MNRAS.382L..30S}
\bibinfo{author}{{Stamatellos}, D.}, \bibinfo{author}{{Hubber}, D.~A.} \& \bibinfo{author}{{Whitworth}, A.~P.}
\newblock \bibinfo{journal}{\bibinfo{title}{{Brown dwarf formation by gravitational fragmentation of massive, extended protostellar discs}}}.
\newblock {\emph{\JournalTitle{\mnras}}} \textbf{\bibinfo{volume}{382}}, \bibinfo{pages}{L30--L34} (\bibinfo{year}{2007}).

\bibitem{2018MNRAS.474.5036F}
\bibinfo{author}{{Forgan}, D.~H.}, \bibinfo{author}{{Hall}, C.}, \bibinfo{author}{{Meru}, F.} \& \bibinfo{author}{{Rice}, W.~K.~M.}
\newblock \bibinfo{journal}{\bibinfo{title}{{Towards a population synthesis model of self-gravitating disc fragmentation and tidal downsizing II: the effect of fragment-fragment interactions}}}.
\newblock {\emph{\JournalTitle{\mnras}}} \textbf{\bibinfo{volume}{474}}, \bibinfo{pages}{5036--5048} (\bibinfo{year}{2018}).

\bibitem{Papaloizou1995}
\bibinfo{author}{{Papaloizou}, J. C.~B.} \& \bibinfo{author}{{Terquem}, C.}
\newblock \bibinfo{journal}{\bibinfo{title}{{On the dynamics of tilted discs around young stars}}}.
\newblock {\emph{\JournalTitle{\mnras}}} \textbf{\bibinfo{volume}{274}}, \bibinfo{pages}{987--1001} (\bibinfo{year}{1995}).

\bibitem{Larwood1996}
\bibinfo{author}{{Larwood}, J.~D.}, \bibinfo{author}{{Nelson}, R.~P.}, \bibinfo{author}{{Papaloizou}, J.~C.~B.} \& \bibinfo{author}{{Terquem}, C.}
\newblock \bibinfo{journal}{\bibinfo{title}{{The tidally induced warping, precession and truncation of accretion discs in binary systems: three-dimensional simulations}}}.
\newblock {\emph{\JournalTitle{\mnras}}} \textbf{\bibinfo{volume}{282}}, \bibinfo{pages}{597--613} (\bibinfo{year}{1996}).

\bibitem{Fragner2010}
\bibinfo{author}{{Fragner}, M.~M.} \& \bibinfo{author}{{Nelson}, R.~P.}
\newblock \bibinfo{journal}{\bibinfo{title}{{Evolution of warped and twisted accretion discs in close binary systems}}}.
\newblock {\emph{\JournalTitle{\aap}}} \textbf{\bibinfo{volume}{511}}, \bibinfo{pages}{A77} (\bibinfo{year}{2010}).

\bibitem{Artymowicz1994}
\bibinfo{author}{{Artymowicz}, P.} \& \bibinfo{author}{{Lubow}, S.~H.}
\newblock \bibinfo{journal}{\bibinfo{title}{{Dynamics of Binary-Disk Interaction. I. Resonances and Disk Gap Sizes}}}.
\newblock {\emph{\JournalTitle{\apj}}} \textbf{\bibinfo{volume}{421}}, \bibinfo{pages}{651} (\bibinfo{year}{1994}).

\bibitem{Pichardo2005}
\bibinfo{author}{{Pichardo}, B.}, \bibinfo{author}{{Sparke}, L.~S.} \& \bibinfo{author}{{Aguilar}, L.~A.}
\newblock \bibinfo{journal}{\bibinfo{title}{{Circumstellar and circumbinary discs in eccentric stellar binaries}}}.
\newblock {\emph{\JournalTitle{\mnras}}} \textbf{\bibinfo{volume}{359}}, \bibinfo{pages}{521--530} (\bibinfo{year}{2005}).

\bibitem{Kley2008}
\bibinfo{author}{{Kley}, W.}, \bibinfo{author}{{Papaloizou}, J.~C.~B.} \& \bibinfo{author}{{Ogilvie}, G.~I.}
\newblock \bibinfo{journal}{\bibinfo{title}{{Simulations of eccentric disks in close binary systems}}}.
\newblock {\emph{\JournalTitle{\aap}}} \textbf{\bibinfo{volume}{487}}, \bibinfo{pages}{671--687} (\bibinfo{year}{2008}).

\bibitem{2007ApJ...654..641J}
\bibinfo{author}{{Jang-Condell}, H.}
\newblock \bibinfo{journal}{\bibinfo{title}{{Constraints on the Formation of the Planet in HD 188753}}}.
\newblock {\emph{\JournalTitle{\apj}}} \textbf{\bibinfo{volume}{654}}, \bibinfo{pages}{641--649} (\bibinfo{year}{2007}).

\bibitem{2006Icar..183..193T}
\bibinfo{author}{{Th{\'e}bault}, P.}, \bibinfo{author}{{Marzari}, F.} \& \bibinfo{author}{{Scholl}, H.}
\newblock \bibinfo{journal}{\bibinfo{title}{{Relative velocities among accreting planetesimals in binary systems: The circumprimary case}}}.
\newblock {\emph{\JournalTitle{\icarus}}} \textbf{\bibinfo{volume}{183}}, \bibinfo{pages}{193--206} (\bibinfo{year}{2006}).

\bibitem{2008MNRAS.388.1528T}
\bibinfo{author}{{Th{\'e}bault}, P.}, \bibinfo{author}{{Marzari}, F.} \& \bibinfo{author}{{Scholl}, H.}
\newblock \bibinfo{journal}{\bibinfo{title}{{Planet formation in {\ensuremath{\alpha}} Centauri A revisited: not so accretion friendly after all}}}.
\newblock {\emph{\JournalTitle{\mnras}}} \textbf{\bibinfo{volume}{388}}, \bibinfo{pages}{1528--1536} (\bibinfo{year}{2008}).

\bibitem{Muller2012}
\bibinfo{author}{{M{\"u}ller}, T.~W.~A.} \& \bibinfo{author}{{Kley}, W.}
\newblock \bibinfo{journal}{\bibinfo{title}{{Circumstellar disks in binary star systems. Models for {\ensuremath{\gamma}} Cephei and {\ensuremath{\alpha}} Centauri}}}.
\newblock {\emph{\JournalTitle{\aap}}} \textbf{\bibinfo{volume}{539}}, \bibinfo{pages}{A18} (\bibinfo{year}{2012}).

\bibitem{2013ApJ...765L...8R}
\bibinfo{author}{{Rafikov}, R.~R.}
\newblock \bibinfo{journal}{\bibinfo{title}{{Planet Formation in Small Separation Binaries: Not so Secularly Excited by the Companion}}}.
\newblock {\emph{\JournalTitle{\apj}}} \textbf{\bibinfo{volume}{765}}, \bibinfo{pages}{L8} (\bibinfo{year}{2013}).

\bibitem{Marzari2009}
\bibinfo{author}{{Marzari}, F.}, \bibinfo{author}{{Scholl}, H.}, \bibinfo{author}{{Th{\'e}bault}, P.} \& \bibinfo{author}{{Baruteau}, C.}
\newblock \bibinfo{journal}{\bibinfo{title}{{On the eccentricity of self-gravitating circumstellar disks in eccentric binary systems}}}.
\newblock {\emph{\JournalTitle{\aap}}} \textbf{\bibinfo{volume}{508}}, \bibinfo{pages}{1493--1502} (\bibinfo{year}{2009}).

\bibitem{2019MNRAS.483.2347D}
\bibinfo{author}{{Desai}, K.~M.}, \bibinfo{author}{{Steiman-Cameron}, T.~Y.}, \bibinfo{author}{{Michael}, S.}, \bibinfo{author}{{Cai}, K.} \& \bibinfo{author}{{Durisen}, R.~H.}
\newblock \bibinfo{journal}{\bibinfo{title}{{A 3D hydrodynamics study of gravitational instabilities in a young circumbinary disc}}}.
\newblock {\emph{\JournalTitle{\mnras}}} \textbf{\bibinfo{volume}{483}}, \bibinfo{pages}{2347--2361} (\bibinfo{year}{2019}).

\bibitem{Kozai1962}
\bibinfo{author}{{Kozai}, Y.}
\newblock \bibinfo{journal}{\bibinfo{title}{{Secular perturbations of asteroids with high inclination and eccentricity}}}.
\newblock {\emph{\JournalTitle{\aj}}} \textbf{\bibinfo{volume}{67}}, \bibinfo{pages}{591--598} (\bibinfo{year}{1962}).

\bibitem{Lidov1962}
\bibinfo{author}{{Lidov}, M.~L.}
\newblock \bibinfo{journal}{\bibinfo{title}{{The evolution of orbits of artificial satellites of planets under the action of gravitational perturbations of external bodies}}}.
\newblock {\emph{\JournalTitle{\planss}}} \textbf{\bibinfo{volume}{9}}, \bibinfo{pages}{719--759} (\bibinfo{year}{1962}).

\bibitem{1986ApJ...309..846L}
\bibinfo{author}{{Lin}, D.~N.~C.} \& \bibinfo{author}{{Papaloizou}, J.}
\newblock \bibinfo{journal}{\bibinfo{title}{{On the Tidal Interaction between Protoplanets and the Protoplanetary Disk. III. Orbital Migration of Protoplanets}}}.
\newblock {\emph{\JournalTitle{\apj}}} \textbf{\bibinfo{volume}{309}}, \bibinfo{pages}{846} (\bibinfo{year}{1986}).

\bibitem{Papaloizou2001}
\bibinfo{author}{{Papaloizou}, J.~C.~B.}, \bibinfo{author}{{Nelson}, R.~P.} \& \bibinfo{author}{{Masset}, F.}
\newblock \bibinfo{journal}{\bibinfo{title}{{Orbital eccentricity growth through disc-companion tidal interaction}}}.
\newblock {\emph{\JournalTitle{\aap}}} \textbf{\bibinfo{volume}{366}}, \bibinfo{pages}{263--275} (\bibinfo{year}{2001}).

\bibitem{Kley2006}
\bibinfo{author}{{Kley}, W.} \& \bibinfo{author}{{Dirksen}, G.}
\newblock \bibinfo{journal}{\bibinfo{title}{{Disk eccentricity and embedded planets}}}.
\newblock {\emph{\JournalTitle{\aap}}} \textbf{\bibinfo{volume}{447}}, \bibinfo{pages}{369--377} (\bibinfo{year}{2006}).

\bibitem{Bitsch2013}
\bibinfo{author}{{Bitsch}, B.}, \bibinfo{author}{{Crida}, A.}, \bibinfo{author}{{Libert}, A.~S.} \& \bibinfo{author}{{Lega}, E.}
\newblock \bibinfo{journal}{\bibinfo{title}{{Highly inclined and eccentric massive planets. I. Planet-disc interactions}}}.
\newblock {\emph{\JournalTitle{\aap}}} \textbf{\bibinfo{volume}{555}}, \bibinfo{pages}{A124} (\bibinfo{year}{2013}).

\bibitem{Ragusa2018}
\bibinfo{author}{{Ragusa}, E.} \emph{et~al.}
\newblock \bibinfo{journal}{\bibinfo{title}{{Eccentricity evolution during planet-disc interaction}}}.
\newblock {\emph{\JournalTitle{\mnras}}} \textbf{\bibinfo{volume}{474}}, \bibinfo{pages}{4460--4476} (\bibinfo{year}{2018}).

\bibitem{2018MNRAS.479.5303L}
\bibinfo{author}{{Lopez}, E.~D.} \& \bibinfo{author}{{Rice}, K.}
\newblock \bibinfo{journal}{\bibinfo{title}{{How formation time-scales affect the period dependence of the transition between rocky super-Earths and gaseous sub-Neptunesand implications for {\ensuremath{\eta}}$_{{\ensuremath{\oplus}}}$}}}.
\newblock {\emph{\JournalTitle{\mnras}}} \textbf{\bibinfo{volume}{479}}, \bibinfo{pages}{5303--5311} (\bibinfo{year}{2018}).

\bibitem{2011ApJS..197....8L}
\bibinfo{author}{{Lissauer}, J.~J.} \emph{et~al.}
\newblock \bibinfo{journal}{\bibinfo{title}{{Architecture and Dynamics of Kepler's Candidate Multiple Transiting Planet Systems}}}.
\newblock {\emph{\JournalTitle{\apjs}}} \textbf{\bibinfo{volume}{197}}, \bibinfo{pages}{8} (\bibinfo{year}{2011}).

\bibitem{2017MNRAS.470.1750I}
\bibinfo{author}{{Izidoro}, A.} \emph{et~al.}
\newblock \bibinfo{journal}{\bibinfo{title}{{Breaking the chains: hot super-Earth systems from migration and disruption of compact resonant chains}}}.
\newblock {\emph{\JournalTitle{\mnras}}} \textbf{\bibinfo{volume}{470}}, \bibinfo{pages}{1750--1770} (\bibinfo{year}{2017}).

\bibitem{2021A&A...650A.152I}
\bibinfo{author}{{Izidoro}, A.} \emph{et~al.}
\newblock \bibinfo{journal}{\bibinfo{title}{{Formation of planetary systems by pebble accretion and migration. Hot super-Earth systems from breaking compact resonant chains}}}.
\newblock {\emph{\JournalTitle{\aap}}} \textbf{\bibinfo{volume}{650}}, \bibinfo{pages}{A152} (\bibinfo{year}{2021}).

\bibitem{2022ApJ...939L..19I}
\bibinfo{author}{{Izidoro}, A.} \emph{et~al.}
\newblock \bibinfo{journal}{\bibinfo{title}{{The Exoplanet Radius Valley from Gas-driven Planet Migration and Breaking of Resonant Chains}}}.
\newblock {\emph{\JournalTitle{\apj}}} \textbf{\bibinfo{volume}{939}}, \bibinfo{pages}{L19} (\bibinfo{year}{2022}).

\bibitem{2023A&A...674A.178B}
\bibinfo{author}{{Bitsch}, B.} \& \bibinfo{author}{{Izidoro}, A.}
\newblock \bibinfo{journal}{\bibinfo{title}{{Giants are bullies: How their growth influences systems of inner sub-Neptunes and super-Earths}}}.
\newblock {\emph{\JournalTitle{\aap}}} \textbf{\bibinfo{volume}{674}}, \bibinfo{pages}{A178} (\bibinfo{year}{2023}).

\bibitem{Owen2013}
\bibinfo{author}{{Owen}, J.~E.} \& \bibinfo{author}{{Wu}, Y.}
\newblock \bibinfo{journal}{\bibinfo{title}{Kepler planets: A tale of evaporation}}.
\newblock {\emph{\JournalTitle{\apj}}} \textbf{\bibinfo{volume}{775}}, \bibinfo{pages}{105} (\bibinfo{year}{2013}).

\bibitem{Ginzburg2018}
\bibinfo{author}{{Ginzburg}, S.}, \bibinfo{author}{{Schlichting}, H.~E.} \& \bibinfo{author}{{Sari}, R.}
\newblock \bibinfo{journal}{\bibinfo{title}{{Core-powered mass-loss and the radius distribution of small exoplanets}}}.
\newblock {\emph{\JournalTitle{\mnras}}} \textbf{\bibinfo{volume}{476}}, \bibinfo{pages}{759--765} (\bibinfo{year}{2018}).

\end{thebibliography}

\end{document}